\documentclass[twocolumn,showpacs,preprintnumbers]{revtex4}
\usepackage{graphicx}
\usepackage{dcolumn}
\usepackage{bm}
\usepackage[bookmarks=false]{hyperref}

\newcommand{\BoldVec}[1]{\mathchoice%
  {\mbox{\boldmath $\displaystyle     #1$}}%
  {\mbox{\boldmath $\textstyle        #1$}}%
  {\mbox{\boldmath $\scriptstyle      #1$}}%
  {\mbox{\boldmath $\scriptscriptstyle#1$}}%
}
\newcommand{\EQ}{\begin{equation}}
\newcommand{\EN}{\end{equation}}
\newcommand{\EQA}{\begin{eqnarray}}
\newcommand{\ENA}{\end{eqnarray}}
\newcommand{\eq}[1]{(\ref{#1})}

\newcommand{\Eq}[1]{Eq.~(\ref{#1})}

\newcommand{\Sec}[1]{\S\,\ref{#1}}

\newcommand{\Fig}[1]{Fig.~\ref{#1}}

\newcommand{\Tab}[1]{Table~\ref{#1}}
\newcommand{\Figs}[2]{Figures~\ref{#1} and \ref{#2}}

\newcommand{\bra}[1]{\langle #1\rangle}

\newcommand{\meanC}{\overline{C}}
\newcommand{\meanF}{\overline{\cal F}}

\newcommand{\meanUU}{\overline{\mbox{\boldmath $U$}}}

\newcommand{\meanFF}{\overline{\mbox{\boldmath ${\cal F}$}} {}}
\newcommand{\zz}{\hat{\mbox{\boldmath $z$}} {}}

\newcommand{\xx}{\BoldVec{x}{}}

\newcommand{\uu}{\BoldVec{u} {}}

\newcommand{\UU}{\BoldVec{U} {}}

\newcommand{\eee}{\BoldVec{e} {}}

\newcommand{\ff}{\BoldVec{f} {}}

\newcommand{\FF}{\BoldVec{F} {}}

\newcommand{\kk}{\BoldVec{k} {}}

\newcommand{\nab}{\BoldVec{\nabla} {}}

\newcommand{\SSSS}{\mbox{\boldmath ${\sf S}$} {}}

\newcommand{\ii}{{\rm i}}

\newcommand{\DD}{{\rm D} {}}
\newcommand{\dd}{{\rm d} {}}

\def\la{\mathrel{\mathchoice {\vcenter{\offinterlineskip\halign{\hfil
$\displaystyle##$\hfil\cr<\cr\sim\cr}}}
{\vcenter{\offinterlineskip\halign{\hfil$\textstyle##$\hfil\cr<\cr\sim\cr}}}
{\vcenter{\offinterlineskip\halign{\hfil$\scriptstyle##$\hfil\cr<\cr\sim\cr}}}
{\vcenter{\offinterlineskip\halign{\hfil$\scriptscriptstyle##$\hfil\cr<\cr\sim\cr}}}}}

\def\half{{\textstyle{1\over2}}}

\def\onethird{{\textstyle{1\over3}}}

\newcommand{\yjas}[4]{, ``#4,'' {\em J.\ Atmos.\ Sci.\ }{\bf #2}, #3 (#1).}

\newcommand{\yana}[4]{, ``#4,'' {\em Astron.\ Astrophys.\ }{\bf #2}, #3 (#1).}

\newcommand{\yjetp}[4]{, ``#4,'' {\em Sov.\ Phys.\ JETP }{\bf #2}, #3 (#1).}

\newcommand{\ymn}[4]{, ``#4,'' {\em Monthly Notices Roy.\ Astron.\ Soc.\ }{\bf #2}, #3 (#1).}

\newcommand{\yjfm}[4]{, ``#4,'' {\em J.\ Fluid Mech.\ }{\bf #2}, #3 (#1).}
\newcommand{\ypr}[4]{, ``#4,'' {\em Phys.\ Rev.\ }{\bf #2}, #3 (#1).}
\newcommand{\yprl}[4]{, ``#4,'' {\em Phys.\ Rev.\ Lett.\ }{\bf #2}, #3 (#1).}

\newcommand{\yapj}[4]{, ``#4,'' {\em Astrophys.\ J.\ }{\bf #2}, #3 (#1).}

\newcommand{\yapjlS}[4]{, ``#4'' {\em Astrophys.\ J.\ Lett.\ }{\bf #2}, #3 (#1).}

\newcommand{\ypf}[4]{, ``#4,'' {\em Phys.\ Fluids }{\bf #2}, #3 (#1).}

\newcommand{\ygafd}[4]{, ``#4,'' {\em Geophys.\ Astrophys.\ Fluid Dynam. }{\bf #2}, #3 (#1).}

\newcommand{\yjour}[5]{, ``#5,'' {\em #2} {\bf #3}, #4 (#1).}

\newcommand{\ybook}[3]{ {\em #2}.\ #3 (#1).}

\begin{document}
\preprint{NORDITA 2003-34 AP}

\title{Non-Fickian diffusion and tau approximation from numerical turbulence}
\author{Axel Brandenburg}
  \email{brandenb@nordita.dk}
  \affiliation{NORDITA, Blegdamsvej 17, DK-2100 Copenhagen \O, Denmark}
\author{Petri J.\ K\"apyl\"a}
  \email{petri.kapyla@oulu.fi}
  \affiliation{Kiepenheuer-Institut f\"ur Sonnenphysik, Sch\"oneckstra\ss{}e 6, D-79104 Freiburg, Germany}
  \affiliation{Department of Physical Sciences, Astronomy Division, P.O. Box 3000, FIN-90014 University of Oulu, Finland}
\author{Amjed Mohammed}
  \email{amjed@mail.uni-oldenburg.de}
  \affiliation{Physics Department, Oldenburg University, 26111 Oldenburg, Germany}
\date{\today,~ $ $Revision: 1.57 $ $}

\begin{abstract}
Evidence for non-Fickian diffusion of a passive scalar is presented using
direct simulations of homogeneous isotropic turbulence.
The results compare favorably with an explicitly time-dependent closure model
based on the tau approximation.
In the numerical experiments three different cases are considered:
(i) zero mean concentration with finite initial concentration flux,
(ii) an initial top hat profile for the concentration, and (iii) an imposed
background concentration gradient.
All cases agree in the resulting relaxation time in the tau approximation
relating the triple correlation to the concentration flux. 
The first order smoothing approximation is shown to be inapplicable.
\end{abstract}
\maketitle

\section{Introduction}

In a turbulent flow the transport of a passive scalar is an important
problem in atmospheric research, astrophysics, and combustion \cite{EKRS00,ES01}.
Passive scalar transport is also an important benchmark for more
complicated turbulent transport processes such as turbulent magnetic
diffusion and the alpha-effect in dynamo theory \cite{Mof78,KR80},
or turbulent viscosity
and its nondiffusive counterparts such as the AKA-effect \cite{SSSF89,BR01}
and the Lambda effect \cite{Rue89,KR93}.

Modeling turbulent transport in terms of turbulent diffusion is known
to have major deficiencies.
For example turbulent transport is known to be anomalous, i.e.\ the
width $\sigma$ of a localized patch of passive scalar concentration may
expand in time like $\sigma^2\sim t^\beta$, where $\beta=1$ corresponds to
ordinary (Brownian) diffusion, $\beta>1$ is superdiffusion, and $\beta<1$
is subdiffusion \cite{BG90}.
Thermal convection, for example, has superdiffusive properties
\cite{MBZ00}.
Turbulent transport is also known to
have nonlocal and nondiffusive properties.
One of the outcomes of this realization was the development of the
transilient matrix approach \cite{Stull84,Stull93} which captures nonlocal
transport properties, although only in a diagnostic fashion \cite{MBZ00}.
In order to describe nonlocal aspects in a prognostic fashion, higher
order spatial derivatives of the turbulent fluxes need to be included.
These are best incorporated in terms of an integral kernel
\cite{BS02}.

In the present work, however, instead of invoking
higher order spatial derivatives, we follow the recent proposal of
Blackman and Field \cite{BF02,BF03}
to include an additional second order {\it time derivative} instead.
This turns the diffusion equation into a damped wave equation.
Blackman and Field derived this equation
from turbulent mean field theory by retaining triple correlations
in the transport equation for the mean flux of a passive scalar.
They assumed an isotropic turbulent flow and use a closure which relates triple
correlations to double correlations \cite{VK83,KRR90,KMR96,RKR03}.
This approach is in some ways more elegant than the classical first
order smoothing approximation \cite{Mof78,KR80,WilliamsFOSA}, which
breaks down because it assumes that
the triple correlations are simply negligible.
This approach also incorporates the momentum equation and,
in magnetohydrodynamics, it therefore allows
a natural derivation of the feedback term
of the alpha effect in magnetohydrodynamics \cite{BF02,RKR03}.

Adding an extra time derivative in the equation for the turbulent transport
of a passive scalar does certainly solve another long
standing problem. Solutions to the diffusion equation are known
to violate causality, because the diffusion equation is elliptic and
the propagation speed of a signal is infinite \cite{ChenLiu94}.
This problem was originally discussed in the context of general relativity
\cite{Israel67}, and more recently in the context of black
hole accretion \cite{KP97,MH98}.
The extra time derivative affects the modeling of turbulent transport most
strongly at early times, just after having injected the passive scalar.
This additional time derivative term tends to make the turbulent transport more ballistic at
early times (corresponding to $\beta\approx2$).
This property is well known in the context of standard Brownian motion.

Non-Fickian diffusion has previously also been discussed in various engineering
applications, for example in diffusion problems in composite media
\cite{ChenLiu03,DepireuxLebon}
and in neutron transport problems in reactors \cite{neutronscattering},
which are best modeled using non-Fickian diffusion.
Here, a non-Fickian
diffusion equation for particle transport arises by taking moments
of the one dimensional Kramers equation, and approximating the
second moment by the Maxwellian value \cite{Das91,ChenLiu03}.
In these applications, however, turbulence is not considered.
One exception is the recent work of Ghosal and Keller \cite{GhosalKeller00}
who derived a non-Fickian diffusion equation with the extra time derivative
by going to the next higher order in an expansion of the underlying
integral equation.
Comparing with data on smoke plumes in the atmosphere and on heat flow
in a wind tunnel they find improved agreement with non-Fickian diffusion
at small distances from the source.

Given that the diffusion equation is now turned into a damped wave
equation, one wonders whether oscillatory behavior is possible.
Blackman and Field \cite{BF03} find that oscillatory behavior
is indeed present for long enough damping times but
disappears for short damping times.
For diffusion of a mean passive scalar, they argue
that the oscillatory behavior is likely unphysical,
and they use this to constrain their damping time to be of order
of the eddy turnover time.
However, the different numerical experiments presented below
suggest that the damping time is about three times longer than
the  eddy turnover time.
Furthermore, the simulations give direct
evidence for mildly oscillatory behavior in a certain parameter regime.

The objective of the present paper is twofold.
First we need to find out whether the existence of the proposed additional
time derivative can actually be confirmed using turbulence simulations.
If so, we need to find out the magnitude of this extra term.
Second, we need to study the range of modifications expected from this
new term.
In order to do this we consider numerical simulations of weakly
compressible turbulence including the transport of a passive scalar.

We begin by discussing the formalism that leads to the emergence of
the additional time derivative in mean field theory.
We then discuss the type of simulations carried out and present three
numerical experiments that quantify the relative importance of the
additional time derivative and that support the tau approximation
formalism.

\section{First order smoothing versus tau approximation}

A classic application of passive scalar transport is the diffusion
of smoke in a turbulent atmosphere.
If the smoke is injected in one point it will diffuse radially
outward, so the mean concentration is expected to be a function
of radius $r$ and time $t$. In that case it makes sense to consider
averages  over spherical shells, i.e.\
\EQ
\meanC(r,t)\equiv{1\over4\pi}\int_0^{2\pi}\int_0^{\pi} C(r,\theta,\phi,t)
\sin\theta\,\dd\theta\,\dd\phi,
\EN
where $C$ is the concentration per unit volume.
Another application is the passive scalar diffusion between
two parts of a slab that are initially separated by a membrane.
In that case the mean concentration  varies along the direction
of the slab, say $z$, and then it makes sense to define horizontal
averages, i.e.\
\EQ
\meanC(z,t)\equiv{1\over L_xL_y}\int_0^{L_x}\int_0^{L_y} C(x,y,z,t)
\,\dd x\,\dd y.
\EN
This is also the type of average that is best suited
for studies in cartesian geometry considered here.

For clarity of the presentation here we ignore
microscopic diffusion, in which case $C$ satisfies
the simple conservation equation,
\EQ
{\partial C\over\partial t}=-\nab\cdot(\UU C),
\label{advection}
\EN
where $\UU$ is the fluid velocity.
The effects of finite microscopic diffusion will be discussed
in the appendix.
We now split $\UU$ and $C$ into mean and fluctuating parts, i.e.\
\EQ
\UU=\meanUU+\uu,\quad C=\meanC+c,
\EN
and average \Eq{advection}, so we have
\EQ
{\partial\meanC\over\partial t}
=-\nab\cdot(\meanUU\,\meanC+\overline{\uu c}).
\label{advection_mean}
\EN
The challenge is now to find an expression for
the concentration flux, $\overline{\uu c}\equiv\meanFF$
in terms of the mean concentration, $\meanC$.
The standard approach is to express the departure of the
concentration from its average, $c\equiv C-\meanC$,
in terms of its past evolution, i.e.\
\EQ
c(\xx,t)=\int_0^t\dot{c}(\xx,t')\,\dd t',
\label{c_past}
\EN
where the dot denotes time differentiation and
\EQ
\dot{c}\equiv\dot{C}-\dot{\meanC}
=-\nab\cdot(\meanUU c+\uu\meanC+\uu c-\overline{\uu c})
\label{cdot}
\EN
is the evolution equation for the passive scalar fluctuation
obtained by subtracting \Eq{advection_mean} from \eq{advection}.
In the first order smoothing approximation or, which is the same, the
quasilinear or second order correlation approximation \cite{Rue89},
one {\it ignores} the terms that are nonlinear in the fluctuations,
i.e.\ the terms $\uu c-\overline{\uu c}$ in \Eq{cdot} are simply omitted
\cite{Mof78,KR80}.
This is only justified if microscopic diffusion is large (but we
have already assumed it to be negligible) or if the velocity is
delta-correlated in time (which is also unrealistic).

The terms that are nonlinear in the fluctuations would lead to
triple correlations of the form $\overline{u_i u_j\partial_j c}$.
Various authors have proposed to approximate triple
correlations by quadratic correlations \cite{VK83,KRR90,KMR96,BF02,RKR03}
which, in the present case, would be $\overline{u_i c}/\tau$; see
Ref.~\cite{BF03}.
This is reminiscent of the Eddy-Damped Quasi-Normal Markovian
approximation \cite{Orszag70,Lesieur90}, where fourth
order correlations are approximated by third order correlations.
This is normally referred to as the tau approximation.
In order to distinguish the two approaches, Blackman and Field \cite{BF03}
call the approach used in Refs~\cite{VK83,KRR90,KMR96,BF02,RKR03} the
``minimal tau approximation''.
In these approaches one calculates not
$\meanFF$, but instead its time derivative.
In that case the time integration in \Eq{cdot} disappears
and one has
\EQ
{\partial\meanFF\over\partial t}
=\overline{\uu(\xx,t)\dot{c}(\xx,t)}
+\overline{\dot{\uu}(\xx,t)c(\xx,t)}.
\label{dmeanFF_dt}
\EN
This leads to the final equation
\EQ
{\partial\meanF_i\over\partial t}=-\overline{u_i u_j}\;\partial_j\meanC
-{\meanF_i\over\tau},
\label{nonFickian}
\EN
where $\tau$ is some relaxation time and incompressibility
has been assumed, i.e.\ $\partial_j u_j=0$.
We shall now also assume isotropy,
$\overline{u_i u_j}=\onethird\delta_{ij} u_{\rm rms}^2$,
where $u_{\rm rms}$
is the rms velocity with $u_{\rm rms}^2=\overline{\uu^2}$.
The validity of \Eq{nonFickian}
is clearly something that ought to be checked numerically
using turbulence simulations.
This is the main objective of the present paper.

The other aspect is that the time derivative may not be
ignorable in the final set of evolution equations.
Thus, in contrast to ordinary Fickian diffusion, where the passive
scalar flux $\meanFF$ is assumed to be proportional to the mean
negative concentration gradient (Fick's law), i.e.\
\EQ
\meanFF=-\kappa_{\rm t}\nab\meanC\quad\mbox{(Fickian diffusion)},
\label{diffapprox}
\EN
where $\kappa_{\rm t}={1\over3}\tau_{\rm cor} u_{\rm rms}^2$ is
the turbulent passive scalar diffusivity and $\tau_{\rm cor}$ is
some correlation time, one now has \cite{BF03}
\EQ
\meanFF=-\kappa_{\rm t}\nab\meanC-\tau{\partial\meanFF\over\partial t}
\quad\mbox{(non-Fickian)},
\label{fluxevol}
\EN
where $\kappa_{\rm t}={1\over3}\tau u_{\rm rms}^2$.
Equation \eq{diffapprox} can be reconciled only when time variations
of the concentration flux have become small and if the correlation
time $\tau_{\rm cor}$ is identified with the damping time $\tau$.

Applying $\partial_t+\tau^{-1}$ on both sides of \eq{advection_mean},
ignoring for simplicity a mean flow ($\meanUU=0$),
and inserting \eq{fluxevol} yields a damped wave equation,
\EQ
{\partial^2\meanC\over\partial t^2}
+{1\over\tau}{\partial\meanC\over\partial t}
=\onethird u_{\rm rms}^2\nabla^2\meanC.
\label{nonFickian_evol}
\EN
We note in passing that the extra term is in some ways analogous to the
displacement current in the Maxwell equations.
This is why this equation is also known in the literature as the
Cattaneo--Maxwell equation \cite{Cattaneo1948}.
The maximum signal speed is limited by $u_{\rm rms}/\sqrt{3}$.
Assessing the importance of the extra time derivative is another
objective of the present paper.

The only ill-known free parameter in this theory is $\tau$,
whose value is conveniently expressed in terms of the Strouhal number
\cite{KR80},
\EQ
\mbox{St}=\tau u_{\rm rms} k_{\rm f},
\label{St_def}
\EN
where $k_{\rm f}$ is the forcing wavenumber or, more generally, the
wavenumber of the scale of the energy carrying eddies.
Here and elsewhere we consider $u_{\rm rms}$ as a constant
(independent of $z$ and $t$).

Some preliminary estimate of $\mbox{St}$ can be made by considering
the late time behavior where Fickian diffusion holds. From \Eq{diffapprox}
we expect that the decay rate of a large scale pattern with
wavenumber $k_1$ is
\EQ
\lambda_c=\kappa_{\rm t}k_1^2,
\EN
where $\kappa_{\rm t}=\onethird\tau u_{\rm rms}^2$ is the turbulent diffusion
coefficient.
From forced turbulence simulations with initial mean flow or mean
magnetic field patterns \cite{YBR03}, the decay rates of
these patterns are well described by a turbulent kinematic viscosity, $\nu_{\rm t}$,
and a turbulent magnetic diffusion coefficient, $\eta_{\rm t}$, where both
coefficients are about equally large with
\EQ
\nu_{\rm t}\approx
\eta_{\rm t}\approx(0.8\ldots0.9)\times u_{\rm rms}/k_{\rm f}.
\EN
Applying the same value also to $\kappa_{\rm t}$ we obtain
\EQ
\mbox{St}\approx (0.8\ldots0.9)\times3=2.4\ldots2.7.
\label{St_estimate}
\EN
This result is remarkable in view of the fact that in the
classic first order smoothing approach to turbulent transport coefficients
one has to assume $\mbox{St}\ll1$; see Refs~\cite{KR80,WilliamsFOSA}.

\section{Comparison with simulations}

In order to test the viability of the non-Fickian diffusion approach
and to determine the value of $\mbox{St}$
we have designed three different types of turbulence simulations.
We first consider the problem of a finite initial flux, $\meanFF$,
but with zero mean concentration, $\meanC=0$ \cite{BF03}.
Next we consider the evolution of an initial top hat profile for $C$
and finally we investigate the case of an imposed uniform gradient
of $C$ which leads to the most direct determination of $\tau$
as a function of Reynolds number and forcing wavenumber.
We begin with a brief description of the simulations carried out.

\subsection{Summary of the type of simulations}

We consider subsonic turbulence in an isothermal
gas with constant sound speed $c_{\rm s}$ in a periodic box of size
$2\pi\times2\pi\times2\pi$.
The Navier--Stokes equation for the velocity $\UU$ is written in the form
\EQ
{\DD\UU\over\DD t}=-c_{\rm s}^2\nab\ln\rho+\FF_{\rm visc}+\ff,
\label{dudt}
\EN
where $\rho$ is the density,
$\DD/\DD t=\partial/\partial t+\UU\cdot\nab$ is the advective
derivative,
\EQ 
\FF_{\rm visc}=\nu\left(\nabla^2\UU+\onethird\nab\nab\cdot\UU
+2\SSSS\cdot\nab\ln\rho\right)
\EN
is the viscous force where 
$\nu=\mbox{const}$ is the kinematic viscosity, 
${\sf S}_{ij}=\frac{1}{2}(U_{i,j}+U_{j,i})-\frac{1}{3}\delta_{ij}U_{k,k}$
is the traceless rate of strain tensor, and $\ff$ is a
random forcing function (see below). The continuity equation is
\EQ
{\partial\rho\over\partial t}=-\nab\cdot(\UU\rho),
\EN
and the equation for the passive scalar concentration per unit volume, $C$, is
\EQ
{\partial C\over\partial t}=-\nab\cdot
\left[\UU C-\rho\kappa_C\nab\left({C\over\rho}\right)\right],
\label{AdvectionDiffusion}
\EN
where $\kappa_C=\mbox{const}$ is the diffusion coefficient for the passive
scalar concentration, which is related to $\nu$ by the Schmidt number,
\EQ
\mbox{Sc}=\nu/\kappa_C.
\EN
Throughout this work we take $\mbox{Sc}=1$.
A nondimensional measure of $\nu$ and hence $\kappa_C$ is the Reynolds
number, which is here defined with respect to the inverse forcing wavenumber,
\EQ
\mbox{Re}=u_{\rm rms}\left/(\nu k_{\rm f})\right..
\label{Rekf}
\EN
The maximum possible value of Re depends on the resolution and
the value of $k_{\rm f}$.
For $k_{\rm f}=1.5$ the typical value is approximately equal to the
number of meshpoints in one direction.

We adopt a forcing function $\ff$ of the form
\EQ
\ff(\xx,t)=\mbox{Re}\{N\ff_{\kk(t)}\exp[i\kk(t)\cdot\xx+i\phi(t)]\},
\EN
where $\xx=(x,y,z)$ is the position vector, and $-\pi<\phi(t)<\pi$ is
a ($\delta$-correlated) random phase.
The normalization factor is
$N=f_0 c_{\rm s}(kc_{\rm s}/\delta t)^{1/2}$, with $f_0$ a
nondimensional forcing amplitude, $k=|\kk|$, and $\delta t$ the length of
the time step;
we chose $f_0=0.05$ so that the maximum Mach number stays below about 0.5
(the rms Mach number is close to 0.2 in all runs \cite{compressibility}).
The vector amplitude $\ff_{\kk}$ describes nonhelical transversal waves with
$|\ff_{\kk}|^2=1$ and
\EQ
\ff_{\kk}=\left(\kk\times\eee\right)/\sqrt{\kk^2-(\kk\cdot\eee)^2},
\EN
where $\eee$ is an arbitrary unit vector.
At each time step we select randomly one of many possible wave vectors
in a finite range around the forcing wavenumber $k_{\rm f}$ (see below).

The equations are solved using the same method as in Ref.~\cite{B01}, but
here we employ a new cache and memory efficient code \cite{PencilCode}
using MPI (Message Passing Interface) library calls for communication between
processors.
This allows us to run at a resolutions up to $1024^3$ meshpoints
\cite{HBD03,DHYB03}.

\subsection{Finite initial flux experiment}
\label{Sinitialflux}

We consider first the example discussed by Blackman and Field \cite{BF03}.
In Fickian diffusion, if $\meanC=0$, there should be no flux, i.e.\
$\meanFF=0$.
Although this should in general be correct, one can imagine contrived
situations where this is not the case, so it is an ideal problem to
test whether the inclusion of the extra time derivative of the flux is
at all correct and meaningful.
Without this extra time derivative $\meanC$ would always stay zero.

To explain in simple terms what happens,
consider a situation where we have initially uniformly mixed white and
black balls (so $\meanC=0$), but for some reason the balls are given
an initial push such that the white balls move to the right part of the domain
and all the black balls move to the left part of the domain.
Then, after a short time, there should be a systematic segregation of
white and black balls, in spite of continuous random forcing.
Of course, this segregation survives only for a dynamical time,
after which ordinary diffusion will begin to mix white and black balls.

In order to set up such a situation in a turbulence simulation we assume
that at $t=0$ the turbulence has already fully developed and then we initialize
the passive scalar distribution according to
\EQ
C(x,y,z,0)=C_0{u_z(x,y,z,0)\over u_{\rm rms}}\sin k_1 z.
\EN
Since $\overline{u}_z=0$, and since the Reynolds rules \cite{KR80} are obeyed by
our horizontal averages, we have $\meanC(z,0)=0$, but
because $\overline{u_z^2}\neq0$, we have $\meanF_z=\overline{u_z c}\neq0$.

Numerically, we monitor the evolution of $\bra{\meanC^2}^{1/2}$,
where angular brackets denote an average over $z$.
This is to be compared with the analytic solution of the model equation
\eq{nonFickian_evol}.
Assuming that $\meanC(z,t)$ is proportional to $\exp(\ii k_1z+\lambda t)$,
the two eigenvalues are 
\EQ
\lambda_\pm(k_1)=-\lambda_0\pm\Delta\lambda(k_1),
\label{disper}
\EN
where
\EQ
\lambda_0={u_{\rm rms} k_{\rm f}\over2\,\mbox{St}},\quad
\Delta\lambda(k_1)=\sqrt{\lambda_0^2-\onethird u_{\rm rms}^2 k_1^2}.
\label{disper2}
\EN
The solution that satisfies $\meanC(z,0)=0$ is
\EQ
\bra{\meanC^2}^{1/2}=A\exp(-\lambda_0 t)\sinh(\Delta\lambda t),
\EN
where $A$ is an amplitude factor.
Oscillatory solutions are possible ($\Delta\lambda$ imaginary)
either when $\mbox{St}$ is large enough or,
since $\mbox{St}$ cannot be manipulated in a simulation, when
$k_{\rm f}$ is small enough.
According to \Eq{St_estimate} we can estimate
\EQ
k_{\rm f}/k_1<2\,\mbox{St}/\sqrt{3}\approx3
\quad\mbox{(oscillatory behavior)}.
\EN
In the oscillatory case, $\Delta\lambda$ is imaginary and so
$\bra{\meanC^2}^{1/2}$ is proportional
to $e^{-\lambda_0 t}|\sin\omega t|$, where $\omega=\mbox{Im}\Delta\lambda$.

\begin{figure}[t!]\begin{center}
\includegraphics[width=.49\textwidth]{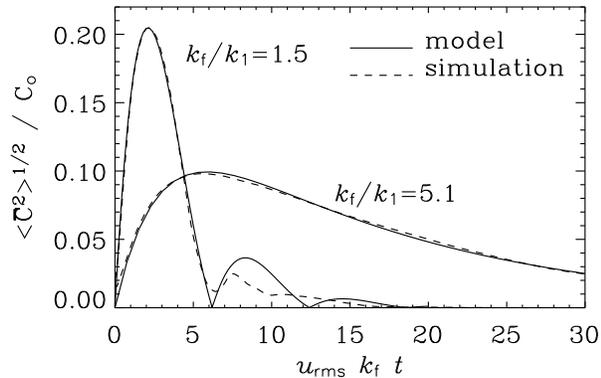}
\end{center}\caption[]{
Passive scalar amplitude, $\bra{\meanC^2}^{1/2}/C_0$,
versus time (normalized by $u_{\rm rms}k_{\rm f}$)
for two different values of $k_{\rm f}/k_1$.
The simulations have $256^3$ meshpoints.
The results are compared with solutions to the
non-Fickian diffusion model.
}\label{plncc_comp}\end{figure}

Note that the solution depends only on the combination
$\mbox{St}/k_{\rm f}$, where $k_{\rm f}$ should be a known
input parameter for a given simulation.
However, in order to be able to fit the model to the simulation we
have considered $\mbox{St}$ and $k_{\rm f}$ as independent fit
parameters and refer then to the quantity $k_{\rm f}^{\rm(fit)}$.
The results of our fits of the simulations to the models are shown in
\Fig{plncc_comp}.
The corresponding fit parameters are listed in \Tab{Tsum}.

\begin{table}[t!]\caption{
Summary of fit parameters for the finite initial flux experiment.
In all cases, the measured value of $u_{\rm rms}=0.23$ is used.
Note that $k_{\rm f}^{\rm(fit)}$ is an independent fit parameter
used instead of $k_{\rm f}$ to model the solution for a given
value of $k_{\rm f}$.
The range of wavenumbers used in the forcing function is also given.
}\vspace{12pt}\centerline{\begin{tabular}{lcccccccccc}
$k_{\rm f}/k_1$  & (range) & $k_{\rm f}^{\rm(fit)}/k_1$ 
& ~~~$\mbox{St}^{\rm(fit)}$~~~ & ~~~$A^{\rm(fit)}$~~~ \\
\hline 
1.5 & (1...2)    & 1.0 & 1.8 & 0.21 \\
2.2 & (2...3)    & 1.6 & 1.8 & 0.38 \\
5.1 & (4.5...5.5)& 3.8 & 2.4 & 0.18 \\
\label{Tsum}\end{tabular}}\end{table}

We see that in all cases the Strouhal number does indeed {\it exceed} unity.
The resulting value is close to the value based on our simple estimate in
\Eq{St_estimate}.
Second, oscillatory behavior of the solution is not only mathematically
possible for small values of $k_{\rm f}$, see \Eq{disper}, but it is even
physically realized in the solution for $k_{\rm f}/k_1=1.5$.

\subsection{Initial top hat function}
\label{Sinitialtophat}

Next we consider the problem of an initial step function.
The advantage of such a profile as initial condition is that
a broad spectrum of wavenumbers is excited.
In order to avoid sharp jumps in the initial condition we choose
a smoothed top hat function using the initial profile
\EQ
C(x,y,z,0)=\half+\half\tanh[k_z^2(d^2-z^2)],
\label{hat_ic}
\EN
where $k_z=2$ and $d=1$ throughout this work.

\begin{figure}[t!]\begin{center}
\includegraphics[width=.49\textwidth]{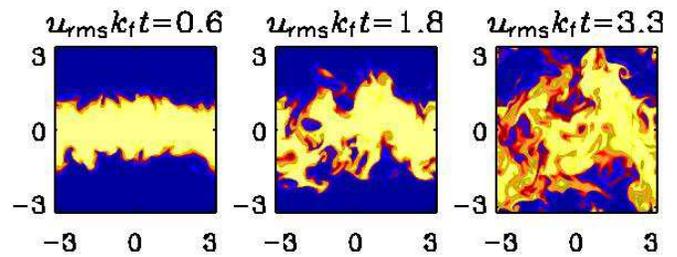}
\end{center}\caption[]{
$C(x,0,z)$  at three different times
after reinitializing $C$ according to \Eq{hat_ic}.
$k_{\rm f}/k_1=1.5$, $\mbox{Re}_{\rm LS}=400$.
}\label{pslices}\end{figure}

It is important to start the experiment at a time
when the turbulence is fully developed.
A visualization of $C$ at three different times
after reinitializing $C$ is shown in \Fig{pslices}.

For Fickian diffusion the initial top hat function will broaden and
develop eventually into a gaussian.
As usual, for large enough values of the Strouhal number, wave-like
behavior is possible and this would correspond to the initial
bump splitting up into two bumps traveling in opposite directions.
We have not been able to see this in our simulations so far.
We have therefore decided to introduce as a quantitative measure of
the departure from a gaussian profile the kurtosis,
\EQ
\kappa={1\over\sigma^4}{\int C z^4\;\dd z\over\int C\;\dd z},
\EN
where $\sigma$ quantifies the width of the profile with
\EQ
\sigma^2={\int C z^2\;\dd z\over\int C\;\dd z}.
\EN
For a gaussian profile we have $\kappa=3$, so we always plot
$\kappa-3$.

At early times, $\sigma^2$ increases quadratically with $t$,
but it soon approaches the linear regime, $\sigma^2\sim t$, until
$\sigma$ saturates at a value comparable to the scale of the box;
see \Eq{nonFickian_evol}.

\begin{figure}[t!]\begin{center}
\includegraphics[width=.49\textwidth]{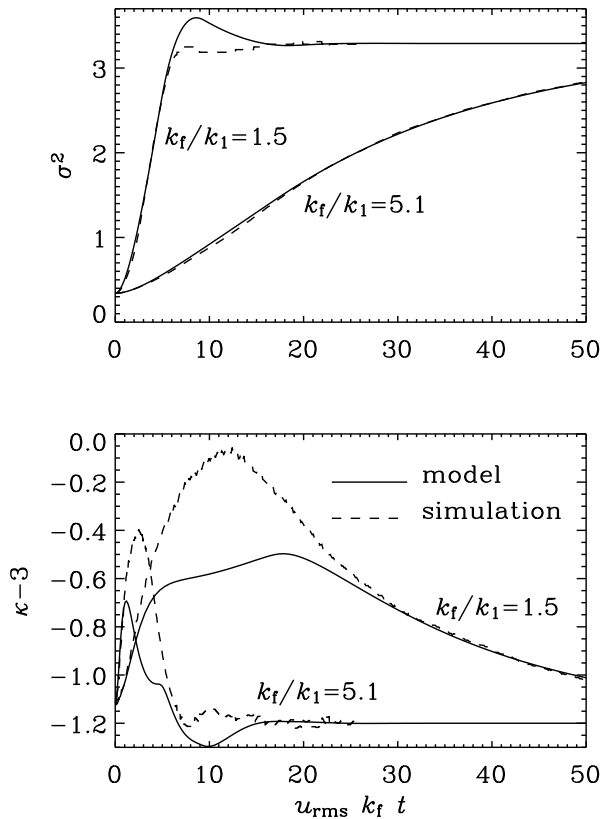}
\end{center}\caption[]{
Comparison of the evolution of $\sigma^2$ and the kurtosis
$\kappa-3$
for the non-Fickian diffusion model and the simulation.
Note the good agreement at early and late times, but there
are departures at intermediate times.
The simulations have $256^3$ meshpoints.
}\label{phat_comp}\end{figure}

In \Fig{phat_comp} we compare the simulation results for $\sigma^2$
and $\kappa-3$ with those obtained from the model \eq{nonFickian_evol}
using the same boundary conditions (periodic in $z$) and for the same
values of $u_{\rm rms}$.
For simplicity we solve \Eq{nonFickian_evol} numerically.
However, similarly to the cases considered in \Sec{Sinitialflux},
we are unable to obtain good fits if we choose exactly the same
values of $k_{\rm f}$ as in the simulation.
Therefore, like in \Sec{Sinitialflux}, we treat $k_{\rm f}$
as a fit parameter denoted by $k_{\rm f}^{\rm(fit)}$; see \Tab{Tsum2}.

\begin{table}[t!]\caption{
Summary of fit parameters for the initial top hat function experiment.
In all cases, the measured value of $u_{\rm rms}=0.23$ is used.
Note that the values of $\mbox{St}^{\rm(fit)}$ are the same as
those used in \Sec{Sinitialflux}, and the values of
$k_{\rm f}^{\rm(fit)}$ are now slightly closer to $k_{\rm f}$
than before.
}\vspace{12pt}\centerline{\begin{tabular}{lcccccccccc}
$k_{\rm f}/k_1$  & $k_{\rm f}^{\rm(fit)}/k_1$ 
& ~~~$\mbox{St}^{\rm(fit)}$~~~ \\
\hline 
1.5 & 1.3 & 1.8 \\
2.2 & 2.0 & 1.8 \\
5.1 & 4.6 & 2.4 \\
\label{Tsum2}\end{tabular}}\end{table}

There are characteristic departures in the values of $\sigma^2$
and $\kappa-3$ for the model compared with the simulations.
This could perhaps be explained by the fact that, especially when
$k_{\rm f}/k_1$ is of order unity, the horizontal averages $\meanC$
obtained from the simulations are strongly `contaminated' by
a small number of large eddies.
Nevertheless, both at early and at late times the agreement
between model and simulation is excellent.

The results in \Sec{Sinitialtophat} confirm our finding of \Sec{Sinitialflux}
that $\mbox{St}$ is around 2 (or even larger).
Again, this is large enough for oscillatory effects to appear when
$k_{\rm f}/k_1$ is small.

\subsection{Imposed mean concentration gradient}

Finally, we consider the case of a uniform gradient in the
mean concentration.
It is advantageous to split $C$ into two contributions,
\EQ
C(x,y,z,t)=\rho(x,y,z,t)Gz+c(x,y,z,t),
\label{ImpGrad}
\EN
where $G=\mbox{const}$ is the imposed mean gradient of the
concentration per unit mass (not unit volume).
Although $C$ is now no longer periodic, this choice still preserves
periodic boundary conditions for the departure $c$ from the background
profile $\rho G z$.
Inserting \Eq{ImpGrad} into \Eq{AdvectionDiffusion} we have
\EQ
{\partial c\over\partial t}=-\nab\cdot
\left[\UU c-\rho\kappa_C\nab\left({c\over\rho}\right)
-\rho\kappa_C G\zz\right]-\rho U_z G,
\EN
where $\zz$ is the unit vector in the $z$ direction.
The main advantage of this setup is the fact that we can now
define mean fields by averaging over the entire volume.
We denote such averages by angular brackets.
Note that $\bra{\UU}=0$, so $\UU=\uu$.
The mean passive scalar flux is then $\bra{\uu c}$
and the triple correlation arising from $\bra{u_z\dot{c}}$ is
\EQ
T_1=\bra{u_z\nab\cdot(\uu c)}.
\EN
Furthermore, there are triple correlation terms arising from the
$\bra{\dot{u}_z c}$ term via the momentum equation.
The $\uu\cdot\nab\uu$ term yields the triple correlation
\EQ
T_2=\bra{(\uu c)\cdot\nab u_z},
\EN
and the pressure gradient term, $\nab p=c_{\rm s}^2\nab\ln\rho$, yields
\EQ
T_3=\bra{c\nabla_{\!z} p},
\EN
where $p=c_{\rm s}^2\ln\rho$ can be regarded as a `reduced' pressure
and is related to the enthalpy.
There is no correlation arising from the forcing term, because the
forcing is delta-correlated in time.
Furthermore, the contributions from the
viscous and diffusive terms are small.
Because of periodic boundary conditions, $T_1+T_2=0$,
so the only contribution surviving from the sum of all three terms
is $T_3$.
Thus, the final expression for $\tau$ is
\EQ
\tau=\bra{u_z c}\left/\bra{c\nabla_{\!z} p}\right..
\EN
We note however that, on the average, the two contributions from
the momentum equations cancel, i.e.\ $T_2+T_3=0$.
Therefore it is also possible to calculate $\tau$ from the
contributions of the passive scalar equation alone, i.e.\
\EQ
\tau=\bra{u_z c}\left/\bra{u_z\nab\cdot(\uu c)}\right..
\EN

\begin{figure}[t!]\begin{center}
\includegraphics[width=.49\textwidth]{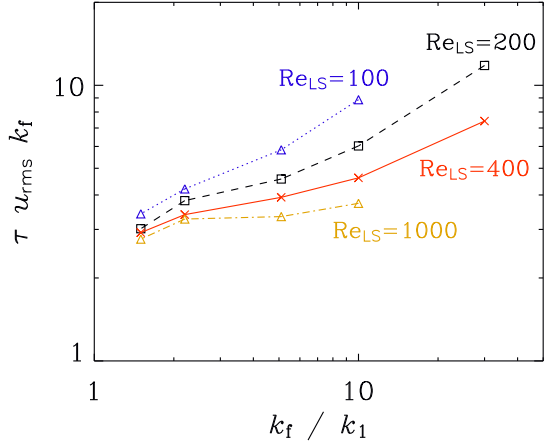}
\end{center}\caption[]{
Strouhal number as a function of $k_{\rm f}/k_1$ for different
values of $\mbox{Re}_{\rm LS}$.
The resolution varies between $64^3$ meshpoints ($\mbox{Re}_{\rm LS}=100$)
and $512^3$ meshpoints ($\mbox{Re}_{\rm LS}=1000$).
}\label{pstrouhal_all_col}\end{figure}

We have calculated a series of simulations for different values
of the Reynolds number as a function of $k_{\rm f}$.
However, for a fixed value of $\nu$, and since $k_{\rm f}$ changes,
the Reynolds number, as defined by \Eq{Rekf}, is not constant.
Therefore we label here the curves by the value of the large
scale Reynolds number that we define as
\EQ
\mbox{Re}_{\rm LS}=u_{\rm rms}\left/(\nu k_1)\right..
\label{ReLS}
\EN
The result is shown in \Fig{pstrouhal_all_col}.

The resulting value of $\mbox{St}$ depends weakly on $k_{\rm f}$ and
increases slowly with increasing $k_{\rm f}$.
This dependence is weaker for smaller values of $k_{\rm f}$.
As the Reynolds number increases, however, the range where
St is approximately constant seems to increase.
It is therefore conceivable that $\mbox{St}$ converges to a universal
constant whose value is around 3.

Comparing with the work of Kleeorin et al.\ \cite{KRR90,KMR96}
one has to note that the $\tau$ approximation was originally formulated
in $k$-space (see also the early work of Orszag \cite{Orszag70}).
In \Eq{nonFickian}, on the other hand, the $\tau$ approximation
is applied directly in real space which may be the reason for
minor differences.
Nevertheless, under the assumption of Kolmogorov turbulence for
$k>k_{\rm f}$, and no turbulence for $k<k_{\rm f}$, one finds that
the Strouhal number is unity.
Given that there can be further discrepancies arising from differences in
the definition of St, we conclude that their result is in broad agreement
with ours.

Since the simulations presented here are weakly compressible,
comparison with incompressible theory may not be quite proper.
If the assumption of incompressibility is relaxed,
i.e.\ $\nab\cdot\uu\neq0$, there is an extra term,
$-\overline{u_i\partial_j u_j}\;\meanC$ on the right hand side of
\Eq{nonFickian}.
In \Eq{nonFickian_evol} this leads to an extra advection term,
$\tau^{-1}\nab\cdot(\meanUU_{\rm eff}\meanC)$ on the left hand side.
Here, $\meanUU_{\rm eff}=\meanUU-\tau\overline{\uu\nab\cdot\uu}$
is a new effective advection velocity; see Refs.~\cite{EKR96,EKR97}.
In the simulations presented here, the term $\overline{\uu\nab\cdot\uu}$
is largest when $k_{\rm f}/k_1$ is small, but even then it is
at most a few percent of $u_{\rm rms}^2k_{\rm f}$.
This justifies {\it a posteriori} the neglect of compressibility effects
in the interpretation of the numerical results.

\begin{figure}[t!]\begin{center}
\includegraphics[width=.49\textwidth]{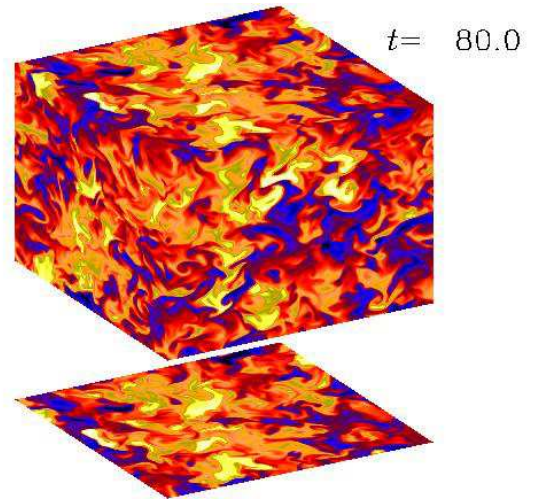}
\end{center}\caption[]{
Visualizations of $C$ on the periphery of the simulation domain
at a time when the simulation has reached a statistically steady state.
$k_{\rm f}/k_1=5.1$, $\mbox{Re}_{\rm LS}=400$.
}\label{nolog256b}\end{figure}

\begin{figure}[t!]\begin{center}
\includegraphics[width=.49\textwidth]{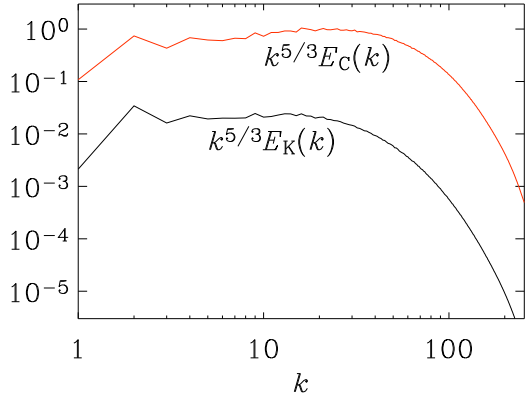}
\end{center}\caption[]{
Compensated kinetic energy and passive scalar spectra for the run with
$k_{\rm f}/k_1=2.2$, $\mbox{Re}_{\rm LS}=1000$.
}\label{power_last}\end{figure}

Visualizations of $C$ on the periphery of the simulation domain
are shown in \Figs{nolog256b}{nolog256a} for $k_{\rm f}=5.1$
and 1.5, respectively \cite{animations}.
Note the combination of large patches (scale $\sim1/k_{\rm f}$)
together with thin filamentary structures.
This is particularly clear in the case with $k_{\rm f}/k_1=1.5$.
The kinetic energy spectrum is close to $k^{-5/3}$, but the
passive scalar spectrum is clearly shallower (perhaps like $k^{1.4}$;
see \Fig{power_last}).
These spectra are, as usual, integrated over shells of constant
$k\equiv|\kk|$ and normalized such that
$\int_0^\infty E_{\rm K}(k)\dd k=\half\bra{\uu^2}$ and
$\int_0^\infty E_{\rm C}(k)\dd k=\half\bra{c^2}$.

\begin{figure}[t!]\begin{center}
\includegraphics[width=.49\textwidth]{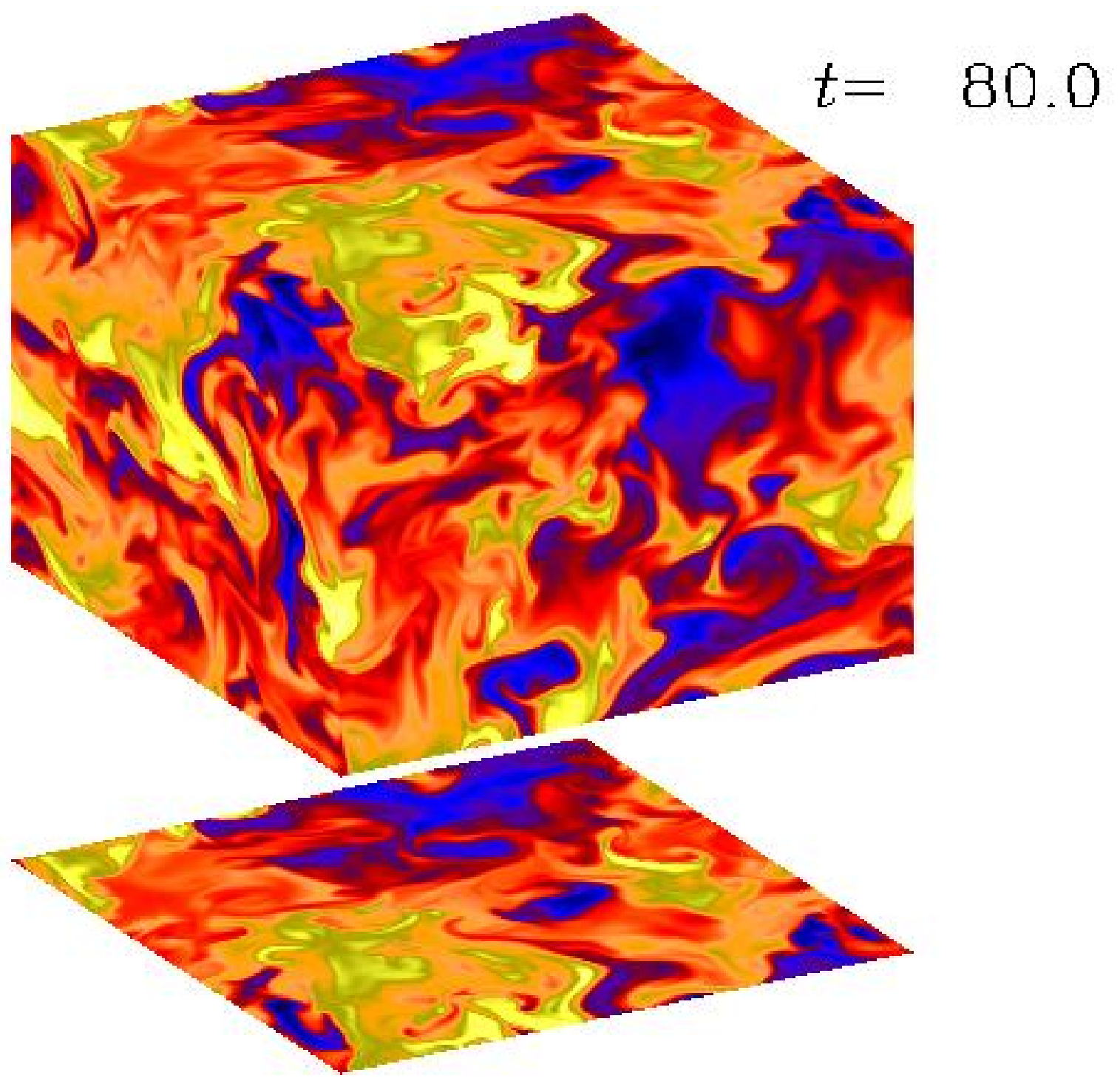}
\end{center}\caption[]{
Same as \Fig{nolog256b}, but for $k_{\rm f}/k_1=1.5$.
}\label{nolog256a}\end{figure}

\section{Conclusions}

Two important results have emerged from the present investigation.
First, the Strouhal number is generally above unity and may have
a universal value between 2 and 3 for forced turbulence.
This implies that the classical first order smoothing approach
in invalid.
Second, the triple correlations that are normally neglected
are of comparable magnitude to the second order corrections
that correspond to the passive scalar flux.
The minimal
tau approximation in which the two are assumed to be proportional
to each other is shown to be justified.

As was shown recently by Blackman and Field in the context of
magnetohydrodynamics \cite{BF02} and then in the context of passive
scalar diffusion \cite{BF03}, this leads to an additional time derivative
in the mean field equation which then takes the form of a damped wave
equation.
Our work has now shown that when the forcing occurs on large enough
scale ($k_{\rm f}\la2k_1$) there is evidence for mildly oscillatory behavior.

Among the various methods for determining the Strouhal number in
a turbulence simulation, the approach of imposing a uniform gradient
of the passive scalar concentration is the most direct one in that
no fitting procedure is needed.
Using this approach requires however the firm knowledge that the functional
form of the mean field equation is correct.
This underlines the importance of the first two approaches where we
were able to compare the evolution of various statistical quantities
with those obtained by solving the model equation.
The only shortcoming here is that we had to find not only
the value of the Strouhal number, but we also had to allow
$k_{\rm f}^{\rm(fit)}$ to deviate (slightly) from the actual value of
$k_{\rm f}$.
Although the difference between the two is perhaps not unreasonable,
one would like to have some theoretical understanding of this discrepancy.

It is remarkable that in all three experiments the value of the Strouhal
number depends only weakly on $k_{\rm f}$.
This suggests that the relaxation time $\tau$ decreases with increasing
values of $k_{\rm f}$; see \Eq{St_def}.
We also emphasize that $\mbox{St}$ is similar in all three experiments,
even though the wavenumber corresponding to the variation of the mean
concentration changed a significantly.
This suggests that $\tau$ does not depend on the scale of the concentration,
even though such a dependence is in principle being allowed for
\cite{Orszag70,KMR96,RKR03}.

The method used in the present paper to determine the Strouhal number from
simulations can straightforwardly be applied to magnetohydrodynamics.
In that case the magnetic field plays the role of the passive scalar
gradient.
Both satisfy very similar equations and in both cases a mean field
can easily be applied while still retaining fully periodic boundary
conditions.
In both cases the closure approach of Blackman and Field predicts
non-Fickian turbulent diffusion and hence the
occurrence of an extra time derivative \cite{BF02,BF03}.
Their analytic approach and
closure agrees reasonably well with our simulations.
Another application would be to determine the role of an extra
time derivative in connection with turbulent viscosity.
In that case a mean gradient could be imposed using the shearing
box approximation \cite{HBG95,BNST95}.
The first two methods described in the present paper should also
still be applicable in that case.
An obvious question that arises in this connection is whether non-Fickian
diffusive properties could also play a role in attempts to find useful
subgrid scale models for Large Eddy Simulations.
The difficulty here is that one has to deal with averages that do
not satisfy the Reynolds rules.
Apart from this difficulty there should be no reason why an extra time
derivative should not also be incorporated in such simulations.

\begin{acknowledgments}
We thank Eric Blackman, Nathan Kleeorin, and G\"unther R\"udiger for
useful comments on our paper.
P.K. acknowledges the financial
support from the Magnus Ehrnrooth foundation and the travel support
from DFG Graduate School `Nonlinear Differential Equations: Modelling,
Theory, Numerics, Visualisation´. 
A.M. acknowledges the financial support from
Hans B\"ockler Stiftung.
P.K. and A.M. wish to thank Nordita and
its staff for their hospitality during their visits.
Use of the parallel computers in Odense (Horseshoe)
and Leicester (Ukaff) is acknowledged.
\end{acknowledgments}



\begin{thebibliography}{99}
\bibitem{compressibility}
For this type of a weakly compressible simulation, we find that the
energies of solenoidal and potential components of the flow have a ratio
$E_{\rm pot}/E_{\rm sol} \approx 10^{-4}\mbox{--}10^{-2}$ for most scales;
only towards the Nyquist frequency the ratio increases to about $0.1$.
It is thus reasonable to assume that compressibility is irrelevant for the
results presented here.
\end{thebibliography}
\end{document}